# Dissecting the Hydrolytic Activities of Sarcoplasmic Reticulum ATPase in the Presence of Acetyl Phosphate*

**Fernando Soler, Maria-Isabel Fortea, Antonio Lax, and Francisco Fernández-Belda‡**

*From the Departamento de Bioquímica y Biología Molecular A, Edificio de Veterinaria, Universidad de Murcia, Campus de Espinardo, 30071 Murcia, Spain*

Sarcoplasmic reticulum vesicles and purified $Ca^{2+}$-ATPase hydrolyze acetyl phosphate both in the presence and absence of $Ca^{2+}$. The $Ca^{2+}$-independent activity was fully sensitive to vanadate, insensitive to thapsigargin, and proceeded without accumulation of phosphorylated enzyme. Acetyl phosphate hydrolysis in the absence of $Ca^{2+}$ was activated by dimethyl sulfoxide. The $Ca^{2+}$-dependent activity was partially sensitive to vanadate, fully sensitive to thapsigargin, and associated with steady phosphoenzyme accumulation. The $Ca^{2+}/P_i$ coupling ratio at neutral pH sustained by 10 mM acetyl phosphate was 0.57. Addition of 30% dimethyl sulfoxide completely blocked $Ca^{2+}$ transport and partially inhibited the hydrolysis rate. Uncoupling induced by dimethyl sulfoxide included the accumulation of vanadate-insensitive phosphorylated enzyme. When acetyl phosphate was the substrate, the hydrolytic pathway was dependent on experimental conditions that might or might not allow net $Ca^{2+}$ transport. The interdependence of both $Ca^{2+}$-dependent and $Ca^{2+}$-independent hydrolytic activities was demonstrated.

The small non-nucleotide substrate acetyl phosphate (AcP)[1] can be hydrolyzed *in vitro* by the SR $Ca^{2+}$-ATPase. When a preparation of native SR vesicles is used, the free energy released from AcP hydrolysis may be partially coupled to the formation of a $Ca^{2+}$ gradient (1–3). Such AcP hydrolysis leads to the steady accumulation of an acid-resistant, hydroxylamine-sensitive EP, as occurs with ATP (4). Other substrates bearing a carboxyl-phosphate anhydride bond such as succinyl phosphate, benzoyl phosphate, carbamyl phosphate (5), and furylacryloyl phosphate (6) can be used by the SR $Ca^{2+}$-ATPase to elicit active $Ca^{2+}$ transport.

Nonetheless, the behavior of AcP as an energy donor substrate is uneven. Other cation-transporting ATPases, such as $H^+,K^+$-ATPase from gastric mucosa (7) or $H^+$-ATPase from yeast plasma membrane (8), are unable to maintain active transport during AcP hydrolysis. This observation has suggested that P-type ATPases do not share the same energy coupling mechanism (7).

* This work was supported by Grant BMC2002-02474 from Spanish Ministerio de Ciencia y Tecnología and Grant PI-22/00756/FS/01 from Fundación Séneca de la Región de Murcia, Spain. The costs of publication of this article were defrayed in part by the payment of page charges. This article must therefore be hereby marked "*advertisement*" in accordance with 18 U.S.C. Section 1734 solely to indicate this fact.
‡ To whom correspondence should be addressed. Tel.: 34-968-364-763; Fax: 34-968-364-147; E-mail: fbelda@um.es.

[1] The abbreviations used are: AcP, acetyl phosphate; SR, sarcoplasmic reticulum; TG, thapsigargin; EP, phosphorylated enzyme intermediate; pNPP, *p*-nitrophenyl phosphate; Mops, 4-morpholinepropanesulfonic acid; Mes, 4-morpholineethanesulfonic acid.

The coupled reaction cycle of SR $Ca^{2+}$-ATPase, as it is usually described (9), involves the participation of phosphorylated and nonphosphorylated enzyme conformations with or without bound $Ca^{2+}$. In fact, conversion of the $Ca^{2+}$-bound phosphorylated conformation into the $Ca^{2+}$-free nonphosphorylated conformation and *vice versa* is the key element in guaranteeing the optimal $Ca^{2+}/P_i$ coupling of 2.

It is also known that SR $Ca^{2+}$-ATPase displays hydrolytic activity on different phosphorylating substrates both in the presence and absence of $Ca^{2+}$ (10–13). The existence of a $Ca^{2+}$-independent activity confirms that the catalytic route may occur exclusively through $Ca^{2+}$-free enzyme conformations. It is self-evident that any hydrolysis occurring through $Ca^{2+}$-free conformations will produce uncoupling. Likewise, it has been shown that an alternative pathway of intramolecular uncoupling may occur through $Ca^{2+}$-bound conformations when phosphorylating substrates, such as ATP (14), UTP (15), or pNPP (12), are hydrolyzed in the presence of $Ca^{2+}$. Uncoupled reaction cycles of the SR $Ca^{2+}$-ATPase have been interpreted as a physiological mechanism of heat production in skeletal muscle fibers (16, 17).

The present study addresses the characterization of hydrolytic activities using AcP as a representative phosphorylating agent bearing a carboxyl-phosphate bond. The steady-state distribution of enzyme conformations with or without bound $Ca^{2+}$ during AcP hydrolysis was evaluated with the aid of the reagents TG, vanadate, and $Me_2SO$. The experimental evidence was completed by assessing whether or not the hydrolytic mechanism included the steady accumulation of EP. This work sheds light on the catalytic and energy transduction mechanism and provides evidence for alternative pathways of substrate utilization by the SR $Ca^{2+}$-ATPase.

## MATERIALS AND METHODS

*Materials*—$[^{45}Ca]CaCl_2$ was a product of PerkinElmer Life Sciences, and potassium $[^{32}P]$phosphate was from Amersham Biosciences. The $Ca^{2+}$ standard solution Titrisol was obtained from Merck. TG was purchased from Molecular Probes Europe, Leiden, The Netherlands. Stock solutions of TG were prepared in $Me_2SO$. Solutions of 1 mM orthovanadate were prepared by dissolving ammonium metavanadate in ultrapure water (Milli-Q grade) adjusted to pH 10.0 with NaOH. The absence of a yellow/orange color confirmed the absence of decavanadate and the presence of monovanadate species (18). AcP (A 0262), $Me_2SO$ (D 8779), deoxycholate (D 4297), and other reagents of analytical grade were obtained from Sigma. Nitrocellulose filter units (HA type) with a 0.45-$\mu$M pore diameter from Millipore and a Hoefer filtration box from Amersham Biosciences were used to evaluate $Ca^{2+}$ transport and EP level.

*SR Vesicles and Purified Enzyme*—A microsomal fraction of sealed vesicles enriched in $Ca^{2+}$-ATPase was obtained from homogenized rabbit skeletal muscle as described by Eletr and Inesi (19). The $Ca^{2+}$-ATPase protein was purified from SR vesicles by partial solubilization with deoxycholate, according to method 2 of Meissner *et al.* (20). Isolated samples were aliquoted and stored at $-80\ °C$ until use. One mg of SR protein contains $\sim$4 nmol of active enzyme, as deduced from the maximal EP level after addition of ATP plus Ca$^{2+}$; therefore, 0.4 mg/ml is equivalent to 1.6 μM Ca$^{2+}$-ATPase.

*AcP Hydrolysis*—Initial rates of AcP hydrolysis were measured at 25 °C, according to Lipmann and Tuttle (21). The colorimetric procedure is based on the evaluation of acetohydroxamic acid as a function of time, which is a measurement of unhydrolyzed AcP. When the enzyme activity was measured at neutral pH and in the presence of 50 μM free Ca$^{2+}$, the reaction medium consisted of 20 mM Mops, pH 7.0, 80 mM KCl, 20 mM MgCl$_2$, 1 mM EGTA, 1.04 mM CaCl$_2$, 5 mM potassium oxalate, 0.4 mg of SR/ml, and 10 mM AcP. Alternatively, the reaction was measured at acidic pH using 20 mM Mes, pH 6.0, as a buffer and decreasing the CaCl$_2$ concentration to 0.608 mM or under alkaline conditions by including 20 mM Tris-HCl, pH 8.0, and 1.05 mM CaCl$_2$. Oxalate and CaCl$_2$ were not added when the experiments were performed in the absence of Ca$^{2+}$. Other conditions were as described in the corresponding figure legends. AcP hydrolysis in the presence or absence of Ca$^{2+}$ was also measured using samples of purified enzyme. In this case, the protein concentration was 0.2 mg/ml, and oxalate was not present.

*Ca$^{2+}$ Transport Experiments*—Initial rates of Ca$^{2+}$ transport were measured at 25 °C with the aid of the radioactive tracer $^{45}$Ca$^{2+}$ (22). The standard reaction medium contained 20 mM Mops, pH 7.0, 80 mM KCl, 20 mM MgCl$_2$, 1 mM EGTA, 1.04 mM [$^{45}$Ca]CaCl$_2$ (~1,500 cpm/nmol), 5 mM potassium oxalate, 0.4 mg of SR/ml, and 10 mM AcP. Aliquots of 0.2-ml reaction mixture (0.08 mg of protein) were manually filtered under vacuum at different time intervals. Filters containing the $^{45}$Ca$^{2+}$-loaded vesicles were rinsed with 10 ml of ice-cold medium consisting of 20 mM Mops, pH 7.0, and 1 mM LaCl$_3$. The radioactivity retained in the filters was measured by liquid scintillation counting.

*Radioactive AcP and Steady-state EP*—[$^{32}$P]AcP was prepared from potassium [$^{32}$P]phosphate and acetic anhydride in a pyridine medium as described by Kornberg *et al.* (23). The reaction medium consisted of 0.25-ml aliquots containing 50 μM free Ca$^{2+}$ (20 mM Mops, pH 7.0, 80 mM KCl, 20 mM MgCl$_2$, 1 mM EGTA, 1.04 mM CaCl$_2$, 5 mM potassium oxalate, and 0.4 mg of SR/ml) or a Ca$^{2+}$-free medium (20 mM Mops, pH 7.0, 80 mM KCl, 20 mM MgCl$_2$, 1 mM EGTA, and 0.4 mg of SR/ml). Phosphorylation at 25 °C was initiated by adding 2 mM [$^{32}$P]AcP (~50,000 cpm/nmol) and allowed to proceed for 30 s when the experiments were performed in the presence of Ca$^{2+}$ or for 1 min when a Ca$^{2+}$-free medium was used. The reaction was stopped by adding 5 ml of ice-cold quenching solution containing 125 mM perchloric acid and 2 mM sodium phosphate. Denatured samples were kept in an ice-water bath for 5 min before manual filtration under vacuum. Filters were extensively rinsed with 50 ml of ice-cold quenching solution and then solubilized and counted by the liquid scintillation technique. The initial reaction medium was supplemented with certain reagents when indicated. A blank assay was performed by adding quenching solution to the sample aliquot before radioactive AcP.

*Other Procedures*—Protein concentration was measured by the procedure of Lowry *et al.* (24) using bovine serum albumin as standard. Free Ca$^{2+}$ was adjusted by the addition of CaCl$_2$ and EGTA stock solutions, as calculated by computation (25). The computer program used the Ca$^{2+}$-EGTA absolute stability constant (26), the H$_4^+$-EGTA dissociation constants (27), and the presence of relevant electrolytes in the medium. For the purpose of this study, the terms absence of Ca$^{2+}$, Ca$^{2+}$-independent, or Ca$^{2+}$-free refer to a low free Ca$^{2+}$ concentration that is insufficient to activate the enzyme.

*Data Presentation*—The plotted mean values correspond to at least three independent assays, each performed in duplicate. The standard errors (plus or minus) are also included. Curve fitting was carried out with the SigmaPlot Graph System from Jandel Scientific.

## RESULTS

The effect of AcP concentration on the hydrolysis rate was initially measured in the presence and in the absence of Ca$^{2+}$. The experimental conditions included native SR vesicles in a buffered medium at neutral pH and the presence of 80 mM K$^+$ and 20 mM Mg$^{2+}$. Free Ca$^{2+}$ was adjusted to 50 μM to measure the Ca$^{2+}$-dependent rate or decreased below the nM range to evaluate the Ca$^{2+}$-independent component. Oxalate was included in measurements of Ca$^{2+}$-dependent activity. A hyperbolic dependence was observed when the hydrolysis rate was plotted as a function of AcP concentration (Fig. 1). The maximal rate calculated from curve fitting was 277 nmol of P$_i$/min/mg of protein in the presence of Ca$^{2+}$ and 138 nmol of P$_i$/min/mg of protein in the absence of Ca$^{2+}$. The $K_m$ value for AcP was 5 mM both in the presence and absence of Ca$^{2+}$, confirming its nature as a low affinity substrate (28).

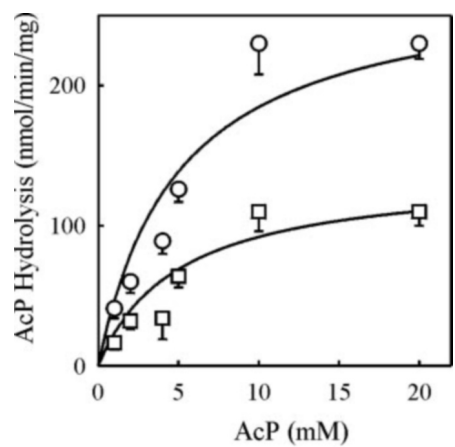

FIG. 1. **Dependence of the hydrolysis rate on AcP concentration using isolated SR vesicles.** Experiments were performed at 25 °C in a medium containing 50 μM free Ca$^{2+}$ (20 mM Mops, pH 7.0, 80 mM KCl, 20 mM MgCl$_2$, 1 mM EGTA, 1.04 mM CaCl$_2$, and 5 mM K$^+$-oxalate) (○) or a Ca$^{2+}$-free medium (20 mM Mops, pH 7.0, 80 mM KCl, 20 mM MgCl$_2$, and 1 mM EGTA) (□). In both cases, the protein concentration was 0.4 mg of SR/ml, and the reaction was started by adding a given AcP concentration.

The AcP hydrolysis rate was also measured at different pH values. The Ca$^{2+}$-dependent activity in SR vesicles increased as pH rose from 6.0 to 7.0 and was very similar at pH 7.0 and 8.0, whereas the Ca$^{2+}$-independent activity showed lower values and was less sensitive to the H$^+$ concentration (Fig. 2*A*). When a purified enzyme preparation was used, the pH dependence of the hydrolytic rate measured in the presence or absence of Ca$^{2+}$ displayed similar behavior (Fig. 2*B*). Measurements of Ca$^{2+}$ transport sustained by 10 mM AcP indicated that the transport rate was higher at neutral pH since it decreased as the pH was lowered to 6.0 or raised to 8.0. Taking into account the corresponding data on hydrolysis in the presence of Ca$^{2+}$, a coupling ratio of 0.57 at neutral pH can be derived. The coupling decreased to 0.31 at pH 6.0 and was close to zero at pH 8.0 (Fig. 2*C*).

The hydrolytic rate in the presence of AcP can be analyzed with the aid of certain reagents. Thus, the sensitivity to TG was studied by measuring enzyme activity at 25 °C and neutral pH using 10 mM AcP as substrate. The Ca$^{2+}$-dependent activity was evaluated in the presence of 50 μM free Ca$^{2+}$ and 5 mM oxalate, whereas the Ca$^{2+}$-independent activity was assayed in the absence of both Ca$^{2+}$ and oxalate. Fig. 3*A* shows that TG produced a concentration-dependent inhibition when the measurements were carried out in a Ca$^{2+}$-containing medium. The hydrolytic activity decreased from 230 nmol of P$_i$/min/mg of protein in the absence of TG to 110 nmol of P$_i$/min/mg of protein when the TG/enzyme molar ratio was ≥1. In contrast, the Ca$^{2+}$-independent activity amounted to ~110 nmol of P$_i$/min/mg of protein and was insensitive to TG even when the inhibitor concentration was raised to 6.4 μM, *i.e.* when the ratio mol of TG/mol of Ca$^{2+}$-ATPase was 4.

The sensitivity to vanadate was also analyzed using the same approach. The inhibition of the Ca$^{2+}$-independent activity by vanadate was consistent with the existence of a single enzyme population, being completely inhibited by ~10 μM vanadate (Fig. 3*B*). However, enzyme activity in the presence of 50 μM free Ca$^{2+}$ displayed a biphasic pattern. A first component, corresponding to 30%, was highly sensitive to vanadate whereas a second component, amounting to 70%, corresponded to a fraction more resistant to inhibition (Fig. 3*B*). Interest-ingly, when the Ca$^{2+}$-containing medium was supplemented with equimolar TG, *i.e.* when TG was 1.6 μM and the SR protein was 0.4 mg/ml, the biphasic dependence became monophasic, and the inhibitory profile in the presence of Ca$^{2+}$ coincided with that observed in the absence of Ca$^{2+}$.

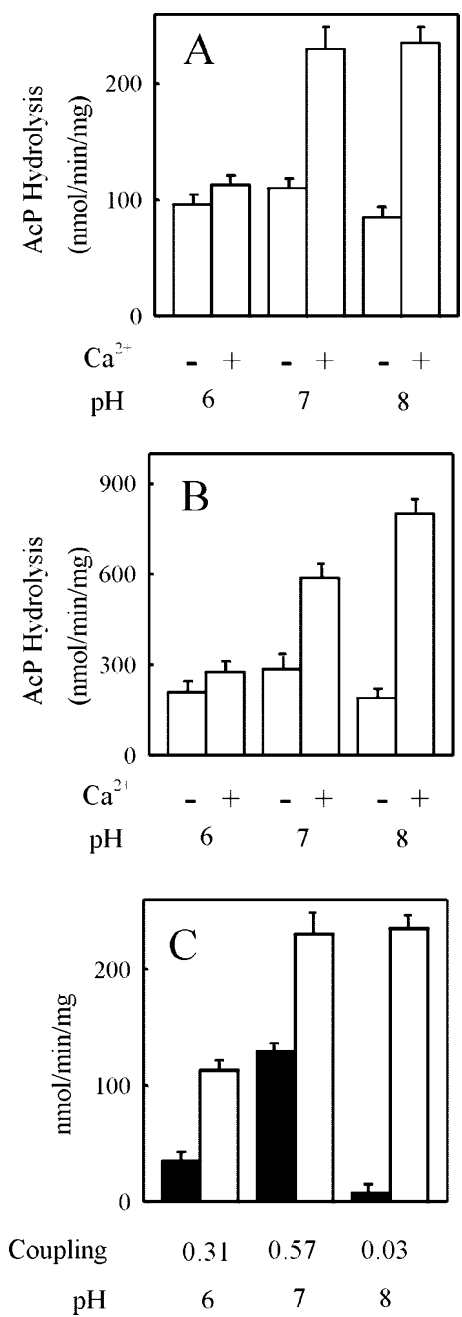

FIG. 2. **Effect of pH on Ca$^{2+}$ activation and Ca$^{2+}$/P$_i$ coupling when AcP was the phosphorylating substrate.** Hydrolytic activities in the absence of Ca$^{2+}$ or in 50 μM free Ca$^{2+}$ medium (*open bars*) and Ca$^{2+}$ transport in the presence of 50 μM free Ca$^{2+}$ (*closed bars*) were measured at pH 6.0, 7.0, or 8.0. The temperature was 25 °C, and 10 mM AcP was the substrate. Experiments were carried out with 0.4 mg/ml SR vesicles (*A* and *C*) or with 0.2 mg/ml purified Ca$^{2+}$-ATPase (*B*). The coupling ratio was obtained by dividing the transport rate by the hydrolysis rate in the presence of Ca$^{2+}$ at the selected pH. The composition of the reaction media is as described under "Materials and Methods."

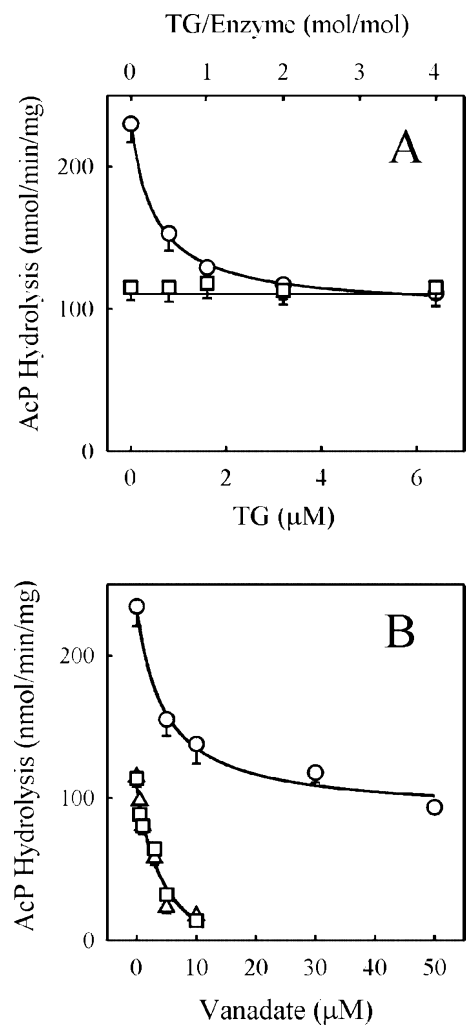

FIG. 3. **Sensitivity to TG or vanadate of Ca$^{2+}$-dependent and Ca$^{2+}$-independent hydrolytic activities in SR vesicles.** The rate of AcP hydrolysis was measured at 25 °C in 50 μM free Ca$^{2+}$ (○) or Ca$^{2+}$-free medium (□). *A*, initial incubation medium as described in the legend of Fig. 1. Then, a given TG concentration was added, and 5 min later, the reaction was started by adding 10 mM AcP. The protein concentration was 0.4 mg/ml SR vesicles, *i.e.* 1.6 μM Ca$^{2+}$-ATPase. In *B*, the composition of reaction medium is as specified in the legend of Fig. 1 but supplemented with a given vanadate concentration. In some experiments, 1.6 μM TG (equimolar) was added during preincubation to the 50 μM free Ca$^{2+}$ medium before the addition of vanadate (△). Reactions were started by adding 10 mM AcP. Fast and slow components in the biphasic dependence on vanadate were evaluated by curve fitting.

Ca$^{2+}$-dependent and Ca$^{2+}$-independent activities displayed different patterns when assayed in the presence of Me$_2$SO. The Ca$^{2+}$-independent activity measured at neutral pH and in the presence of 10 mM AcP was linearly activated from 110 to 270 nmol of P$_i$/min/mg of protein when the Me$_2$SO concentration was raised from 0 to 30% (v/v) (Fig. 4*A*). In contrast, the enzyme activity in the presence of 50 μM free Ca$^{2+}$ was partially inhibited when the organic solvent was raised in the same concentration range. The rate of AcP hydrolysis in the presence of Ca$^{2+}$ was 230 nmol of P$_i$/min/mg of protein in the absence of organic solvent and 118 nmol of P$_i$/min/mg of protein when 30% Me$_2$SO was present.

The functional effect of Me$_2$SO was also studied by measuring Ca$^{2+}$ transport in a medium containing 50 μM free Ca$^{2+}$ and 10 mM AcP. The Ca$^{2+}$ transport rate at neutral pH was 130 nmol of Ca$^{2+}$/min/mg of protein when measured in the absence of Me$_2$SO but practically zero when measured in the presence of 30% Me$_2$SO (Fig. 4*B*). The effect of Me$_2$SO on SR vesicle permeability was tested previously by adding the organic solvent once the active transport process was initiated. The addition of 30% Me$_2$SO after 9 min of reaction did not alter the Ca$^{2+}$ content already accumulated inside the vesicles (data not shown), thus ruling out any ionophoric activity.

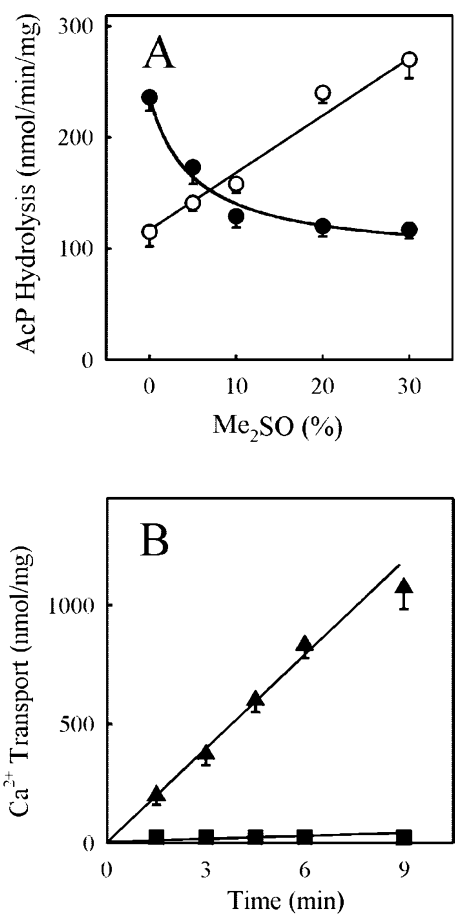

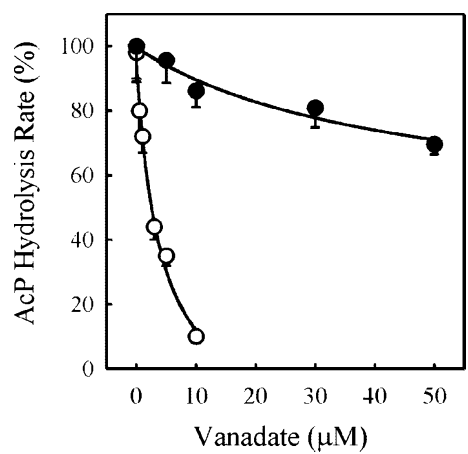

FIG. 5. **Sensitivity to vanadate of hydrolytic activities measured in the presence of 30% $Me_2SO$.** Hydrolysis of AcP by SR vesicles was measured at 25 °C and neutral pH in the 50 μM free $Ca^{2+}$ (●) or $Ca^{2+}$-free medium (○) supplemented with 30% $Me_2SO$ (v/v). The inhibitory effect of vanadate was studied by including different vanadate concentrations in the reaction medium.

FIG. 4. **Effect of $Me_2SO$ on AcP hydrolysis and $Ca^{2+}$ transport.** In *A*, SR vesicles (0.4 mg/ml) were initially equilibrated at neutral pH in 50 μM free $Ca^{2+}$ (●) or $Ca^{2+}$-free medium (○). Then, a given $Me_2SO$ concentration (v/v) was added, and the rate of AcP hydrolysis was measured at 25 °C in the presence of 10 mM AcP. The composition of reaction media is as described in the legend of Fig. 1. In *B*, the time course of $Ca^{2+}$ transport was measured at 25 °C in a medium containing 20 mM Mops, pH 7.0, 80 mM KCl, 20 mM $MgCl_2$, 1 mM EGTA, 1.04 mM $^{45}Ca$-$Cl_2$ (50 μM free $Ca^{2+}$), 0.4 mg/ml SR vesicles, 5 mM $K^+$-oxalate, and 10 mM AcP in the absence (▲) or presence of 30% $Me_2SO$ (v/v) (■).

The uncoupling process induced by $Me_2SO$ was characterized by studying the sensitivity to vanadate. To this end, the experiments shown in Fig. 3*B* were now repeated in the presence of $Me_2SO$. When SR vesicles in a $Ca^{2+}$-free medium were supplemented with 30% $Me_2SO$ and 10 mM AcP was present, the hydrolysis rate was highly sensitive to vanadate inhibition, as observed in the absence of organic solvent (*cf.* Fig. 5 and Fig. 3*B*). However, the enzyme activity in the presence of 50 μM free $Ca^{2+}$ and 30% $Me_2SO$ was hardly sensitive to vanadate (Fig. 5). The sensitivity of the $Ca^{2+}$-dependent activity to vanadate was lower in the presence than in the absence of organic solvent (*cf.* Fig. 5 and Fig. 3*B*).

Steady accumulation of radioactive EP under turnover conditions was evaluated by adding [$^{32}P$]AcP (Fig. 6). Maximal EP levels were observed when SR vesicles were phosphorylated in the standard 50 μM free $Ca^{2+}$ medium. EP accumulation in the $Ca^{2+}$-containing medium was practically abolished by equimolar TG but was almost totally insensitive to vanadate. Furthermore, AcP hydrolysis in the $Ca^{2+}$-containing medium and in the presence of 30% $Me_2SO$ was associated with partial accumulation of vanadate-insensitive EP. No EP was accumulated when a $Ca^{2+}$-free medium was used, and thus, TG or vanadate had no effect under this condition. AcP hydrolysis in the absence of $Ca^{2+}$ but in the presence of 30% $Me_2SO$ led to practi-cally full enzyme phosphorylation that was, in this case, sensitive to vanadate.

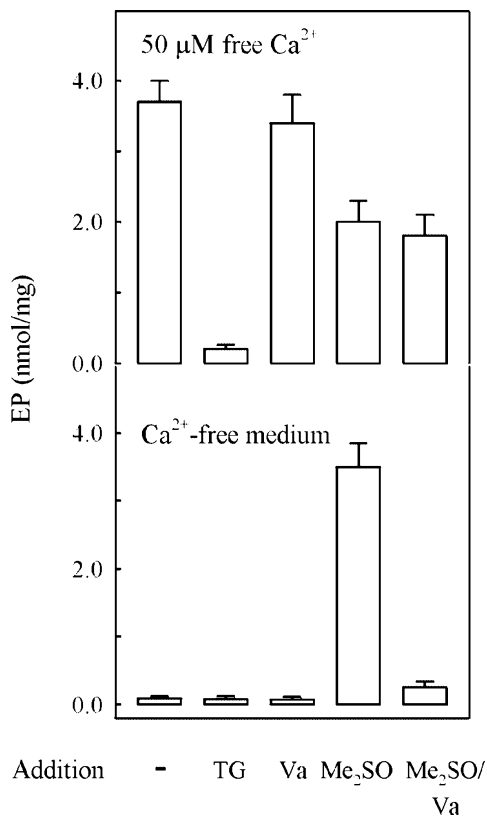

FIG. 6. **EP accumulation sustained by AcP during the enzyme turnover.** Experiments were performed at 25 °C in 50 μM free $Ca^{2+}$ (*upper row*) or $Ca^{2+}$-free medium (*lower row*) as described under "Materials and Methods." The phosphorylating substrate was 2 mM [$^{32}P$]AcP, and the phosphorylation time was 30 s for the $Ca^{2+}$-containing samples or 1 min when a $Ca^{2+}$-free medium was used. The reaction medium was supplemented with 1.6 μM TG (TG/enzyme = 1), 10 μM vanadate (*Va*), 30% $Me_2SO$, or 30% $Me_2SO$ plus 10 μM vanadate when indicated.

DISCUSSION

Isolated SR vesicles as well as purified $Ca^{2+}$-ATPase displayed AcP hydrolytic activity when measured in the presence or absence of $Ca^{2+}$ (Fig. 2, *A* and *B*). This observation was already made with the substrates ATP (13) and pNPP (12). Both activities showed the same $K_m$ values for AcP, and the maximal hydrolytic rate in the presence of $Ca^{2+}$ was only twice that observed in the absence of $Ca^{2+}$ (Fig. 1). These features clearly indicated that both activities were sustained by the $Ca^{2+}$-ATPase protein. Previous studies using SR vesicles had suggested that the $Ca^{2+}$-independent activity was linked to a contaminating phosphatase activity (29).

A major difference was the steady accumulation of EP when the hydrolysis occurred in a $Ca^{2+}$-containing medium as opposed to the lack of EP when a $Ca^{2+}$-free medium was used (Fig. 6). It seems that AcP can gain access to the enzyme catalytic site either in the presence or in the absence of $Ca^{2+}$, and therefore, $Ca^{2+}$ binding is not a prerequisite for AcP hydrolysis. Nevertheless, the hydrolytic process is more efficient when it occurs in a $Ca^{2+}$-containing medium. The phosphorylation rate in the presence of AcP plus $Ca^{2+}$ is quite slow when compared with the rate in the presence of ATP plus $Ca^{2+}$ although sufficiently faster than the dephosphorylation rate to allow EP accumulation (29). However, the hydrolysis rate, and presumably the phosphoryl transfer reaction, are slower when the reaction takes place in the absence of $Ca^{2+}$ and do not compensate for EP cleavage. For this reason, no EP accumulation is usually observed in a $Ca^{2+}$-free medium.

According to the conventional reaction cycle, the steady accumulation of EP is associated with $Ca^{2+}$-bound conformations as opposed to uncoupled hydrolysis occurring through $Ca^{2+}$-free species. However, this is not always the case. Thus, hydrolysis of pNPP by SR $Ca^{2+}$-ATPase in the presence of $Ca^{2+}$ does not allow EP accumulation unless the experimental conditions are forced (30). Moreover, furylacryloyl phosphate hydrolysis in the absence of $Ca^{2+}$ but in the presence of 30% $Me_2SO$ produces an accumulation of ~2 nmol of EP/mg of protein (6), and EP is accumulated when the $Na^+,K^+$-ATPase is in the presence of ATP, $K^+$, and $Me_2SO$ but in the absence of $Na^+$ (31). Furthermore, ATP hydrolysis by the plasma membrane $Ca^{2+}$-ATPase in the presence of $Ca^{2+}$ produces low EP levels when compared with the maximal value (32). Therefore, EP accumulation is only indicative of enzyme turnover through $Ca^{2+}$-bound conformations in certain conditions since it is affected by parameters such as nature of the substrate, reaction temperature, free $Ca^{2+}$ inside the vesicles, presence of organic solvent, etc.

The rate of AcP hydrolysis was the same in the presence or absence of $Ca^{2+}$ when ≥1 mol of TG/mol of enzyme was added (Fig. 3A). This observation can be explained by enzyme activity interconversion since: (i) TG stabilizes the enzyme in the $Ca^{2+}$-free conformation (33, 34), (ii) TG does not inhibit the $Ca^{2+}$-independent activity (13), and (iii) both hydrolytic activities are derived from the same protein. In other words, the enzyme in a $Ca^{2+}$-containing medium is forced by TG to express the $Ca^{2+}$-independent activity. This is also suggested by the fact that no EP was accumulated in the presence of $Ca^{2+}$ when TG was added, as occurs during the enzyme turnover in a $Ca^{2+}$-free medium (Fig. 6).

The existence of a major vanadate-resistant component, which is evident when AcP hydrolysis is measured at neutral pH and in a $Ca^{2+}$-containing medium (Fig. 3B), suggests the prevalent accumulation of $Ca^{2+}$-bound conformations since vanadate inhibits the $Ca^{2+}$-independent activity. $Ca^{2+}$-bound species were tested by repeating experiments in the presence of equimolar TG. The inhibitor TG blocked the whole enzyme population in the $Ca^{2+}$-free conformation (Fig. 3A), and the vanadate-dependent inhibitory profile measured in the presence of $Ca^{2+}$ plus TG exactly matched that observed in the absence of $Ca^{2+}$ (Fig. 3B). The fact that EP reached almost maximal levels in the presence of $Ca^{2+}$ when vanadate was added (Fig. 6) confirms that $Ca^{2+}$-bound conformations were involved in the prevalent hydrolytic pathway.

$Me_2SO$ produced opposite effects on AcP hydrolysis rates depending on the presence or absence of $Ca^{2+}$ (Fig. 4A). Namely, the rate of AcP hydrolysis in the presence of $Ca^{2+}$ was partially inhibited when 30% $Me_2SO$ was present. Also, AcP hydrolysis in the presence of $Ca^{2+}$ plus 30% $Me_2SO$ did not sustain net $Ca^{2+}$ transport (Fig. 4B), giving rise to energy uncoupling. Our data indicate that the absence of $Ca^{2+}$ transport induced by 30% $Me_2SO$ was associated with vanadate-resistant species (Fig. 5). Additional evidence was the steady accumulation of vanadate-resistant EP (Fig. 6). The absence of $Ca^{2+}$ transport in the presence of $Me_2SO$ with AcP as substrate was attributed previously to energy uncoupling through $Ca^{2+}$-free conformations (11, 27). In this regard, 40% $Me_2SO$ favored the accumulation of vanadate-sensitive species, *i.e.* $Ca^{2+}$-free conformations when the substrate was pNPP and $Ca^{2+}$ was present (12).

This study reveals that hydrolysis and uncoupling mainly occurred through $Ca^{2+}$-bound conformations and steady EP accumulation as can be observed with the substrates ATP or pNPP. Hydrolysis and uncoupling in the presence of $Ca^{2+}$ and $Me_2SO$ also occurred mainly through $Ca^{2+}$-bound conformations and EP species when the substrate was AcP, at variance with the data obtained with the substrate pNPP (12).

The $Ca^{2+}$-independent activity in this study is mechanistically similar to the $Na^+,K^+$-ATPase activity when AcP or pNPP is hydrolyzed in the presence of $K^+$ and absence of $Na^+$. The so-called phosphatase activity does not support cation transport and has been attributed to $E_2$ conformations (35, 36). The present results highlight the interdependence of $Ca^{2+}$-dependent and $Ca^{2+}$-independent hydrolytic activities catalyzed by SR $Ca^{2+}$-ATPase, and therefore, the versatility of the enzyme reaction cycle.

REFERENCES

1. de Meis, L. (1969) *J. Biol. Chem.* **244**, 3733–3739
2. de Meis, L., and Hasselbach, W. (1971) *J. Biol. Chem.* **246**, 4759–4763
3. Pucell, A., and Martonosi, A. (1971) *J. Biol. Chem.* **246**, 3389–3397
4. Friedman, Z., and Makinose, M. (1970) *FEBS Lett.* **11**, 69–72
5. Liguri, G., Stefani, M., Berti, A., Nassi, P., and Ramponi, G. (1980) *Arch. Biochem. Biophys.* **200**, 357–363
6. Inesi, G., Kurzmack, M., Nakamoto, R., de Meis, L., and Bernhard, S. A. (1980) *J. Biol. Chem.* **255**, 6040–6043
7. Asano, S., Kamiya, S., and Takeguchi, N. (1992) *J. Biol. Chem.* **267**, 6590–6595
8. Wang, G., and Perlin, D. S. (1977) *Arch. Biochem. Biophys.* **344**, 309–315
9. de Meis, L., and Vianna, A. L. (1979) *Annu. Rev. Biochem.* **48**, 275–292
10. Carvalho-Alves, P. C., and Scofano, H. M. (1987) *J. Biol. Chem.* **262**, 6610–6614
11. Chini, E. N., Montero-Lomeli, M., and de Meis, L. (1990) *Biochim. Biophys. Acta* **1030**, 152–156
12. Fernandez-Belda, F., Fortea, M. I., and Soler, F. (2001) *J. Biol. Chem.* **276**, 7998–8004
13. Fortea, M. I., Soler, F., and Fernandez-Belda, F. (2001) *J. Biol. Chem.* **276**, 37266–37272
14. Yu, X., and Inesi, G. (1995) *J. Biol. Chem.* **270**, 4361–4367
15. Fortea, M. I., Soler, F., and Fernandez-Belda, F. (2000) *J. Biol. Chem.* **275**, 12521–12529
16. Lowell, B. B., and Spiegelman, B. M. (2000) *Nature* **404**, 652–660
17. de Meis, L. (2001) *J. Biol. Chem.* **276**, 25078–25087
18. Hua, S., Fabris, D., and Inesi, G. (1999) *Biophys. J.* **77**, 2217–2225
19. Eletr, S., and Inesi, G. (1972) *Biochim. Biophys. Acta* **282**, 174–179
20. Meissner, G., Conner, G. E., and Fleischer, S. (1973) *Biochim. Biophys. Acta* **298**, 246–269
21. Lipmann, F., and Tuttle, L. (1945) *J. Biol. Chem.* **159**, 21–28
22. Martonosi, A., and Feretos, R. (1964) *J. Biol. Chem.* **239**, 648–658
23. Kornberg, A., Kornberg, S. R., and Simms, E. S. (1956) *Biochim. Biophys. Acta* **20**, 215–227
24. Lowry, O. H., Rosebrough, N. J., Farr, A. L., and Randall, R. J. (1951) *J. Biol. Chem.* **193**, 265–275
25. Fabiato, A. (1988) *Methods Enzymol.* **157**, 378–417
26. Schwartzenbach, G., Senn, H., and Anderegg, G. (1957) *Helv. Chim. Acta* **40**, 1886–2000
27. Blinks, J., Wier, W., Hess, P., and Prendergast, F. (1982) *Prog. Biophys. Mol. Biol.* **40**, 1–114
28. Bodley, A. L., and Jencks, W. P. (1987) *J. Biol. Chem.* **262**, 13997–14004

29. Teruel, J. A., and Inesi, G. (1988) *Biochemistry* **27,** 5885–5890
30. Nakamura, Y., and Tonomura, Y. (1978) *J. Biochem.* (*Tokyo*) **83,** 571–583
31. Barrabin, H., Fontes, C. F., Scofano, H. M., and Norby J. G. (1990) *Biochim. Biophys. Acta* **1023,** 266–273
32. Kosk-Kosicka, D., Scaillet, S., and Inesi, G. (1986) *J. Biol. Chem.* **261,** 3333–3338
33. Sagara, Y., and Inesi, G. (1990) *J. Biol. Chem*. **266,** 13503–13506
34. Sagara, Y., Fernandez-Belda, F., de Meis, L., and Inesi, G. (1992) *J. Biol. Chem.* **267,** 12606–12613
35. Robinson, J. D., Levine, G. M., and Robinson, L. J. (1983) *Biochim. Biophys. Acta* **731,** 406–414
36. Berberian, G., and Beauge, L. (1985) *Biochim. Biophys. Acta* **821,** 17–29